\documentclass[pre,amsfonts,amssymb,amsmath,twocolumn,floatfix]{revtex4}
\usepackage{natbib}
\usepackage{graphicx}

\newcommand{\etal}{\textit{et al.}}

\newcommand{\vc}[1]{\mathbf{#1}}

\newcommand{\abs}[1]{\left|#1\right|}
\newcommand{\ept}[1]{\left\langle#1\right\rangle}

\DeclareMathOperator{\Var}{Var}

\begin{document}
\title{Dynamics of Random Packings in Granular Flow}
\author{Chris H. Rycroft}
\email{chr@mit.edu}
\author{Martin Z. Bazant}
\email{bazant@mit.edu}
\affiliation{Department of Mathematics, Massachusetts Institute of Technology, Cambridge, MA 02139}
\author{Gary S. Grest}
\affiliation{Sandia National Laboratories, Albuquerque, NM 87185}
\author{James W. Landry}
\affiliation{Lincoln Laboratory, Massachusetts Institute of Technology, Lexington, MA 02420}
\begin{abstract}
We present a multiscale simulation algorithm for amorphous materials, which we
illustrate and validate in a canonical case of dense granular flow. Our
algorithm is based on the recently proposed Spot Model, where particles in a
dense random packing undergo chain-like collective displacements in response to
diffusing ``spots'' of influence, carrying a slight excess of interstitial free
volume. We reconstruct the microscopic dynamics of particles from the ``coarse
grained'' dynamics of spots by introducing a localized particle relaxation step
after each spot-induced block displacement, simply to enforce packing
constraints with a (fairly arbitrary) soft-core repulsion. To test the model,
we study to what extent it can describe the dynamics of up to 135,000
frictional, viscoelastic spheres in granular drainage simulated by the
discrete-element method (DEM). With only five fitting parameters (the radius,
volume, diffusivity, drift velocity, and injection rate of spots), we find that
the spot simulations are able to largely reproduce not only the mean flow and
diffusion, but also some subtle statistics of the flowing packings, such as
spatial velocity correlations and many-body structural correlations. The spot
simulations run over 100 times faster than DEM and demonstrate the possibility
of multiscale modeling for amorphous materials, whenever a suitable model can
be devised for the coarse-grained spot dynamics.
\end{abstract}
\maketitle

\section{Introduction}\label{sec:intro}
\begin{figure}
(a)\includegraphics[width=1.3in]{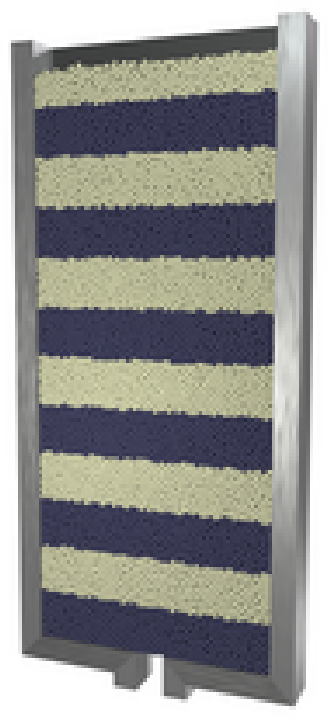} 
(b)\includegraphics[width=1.3in]{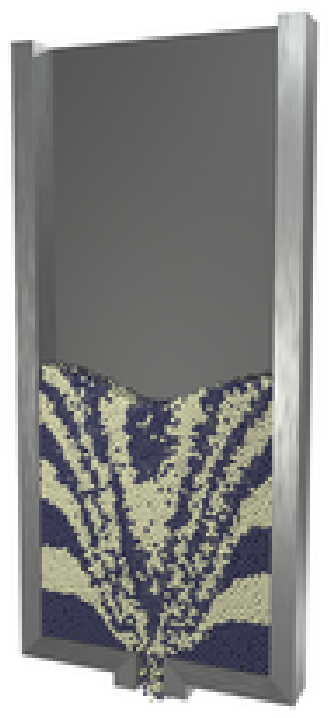}
\caption{(Color online) A simulation of the experiment in
Ref.~\protect\cite{choi04} by discrete element simulations. (a) First, 55,000
glass beads are poured into a quasi-two-dimensional silo (8 beads deep) and let
come to rest. (b) Slow drainage occurs after a slit orifice is opened. (The
grains are identical, but colored by their initial height.)
\label{fig:initial} }
\end{figure}

The geometry of static sphere packings is an age-old problem~\cite{torquato}
with current work focusing on jammed random packings~\cite{torquato00,ohern03},
but how do random packings flow? Here, we consider the case of granular
drainage~\cite{jaeger96}, which is of practical importance (e.g. in pebble-bed
nuclear reactors~\cite{talbot02,reactor}) and also raises fundamental questions
in non-equilibrium statistical mechanics~\cite{kadanoff99}. In fast, dilute
flows, Boltzmann's kinetic theory of gases can be modified to account for
inelastic collisions~\cite{jenkins83}, but slow, dense flows (as in
Fig.~\ref{fig:initial}) require a different description due to long-lasting,
many-body contacts~\cite{choi04}.  Although ballistic motion may occur at the
nano-scale~\cite{menon97} ($< 0.01$\% of a grain diameter), collisions do not
result in random recoils, as in a gas.

In crystals, diffusion and flow are mediated by defects, such as vacancies and
dislocations, but in disordered phases it is not clear what, if any,
``defects'' might facilitate structural rearrangements. Perhaps the only
candidate in the literature is an empty ``void'' in the random packing into
which a single particle may hop, thereby displacing the void. The void
mechanism was proposed by Eyring for viscous flow~\cite{eyring36} and has
re-appeared in theories of the glass transition~\cite{cohen59}, shear flow in
metallic glasses~\cite{spaepen77}, compaction in vibrated granular
materials~\cite{boutreux97}, and granular drainage from a
silo~\cite{mullins72}, but it is now seen as unrealistic. In glasses,
cooperative relaxation (involving many particles at once) has been
observed~\cite{glotzer98,weeks00}, presumably facilitated by free
volume~\cite{cohen79,lemaitre02,garrahan03}. In granular drainage, the Void
Model gives a reasonable fit to the mean flow~\cite{tuzun79,choiexpt}, and yet
it grossly over-predicts diffusion~\cite{choi04}.

\begin{figure*}
(a)\includegraphics[width=0.25\textwidth]{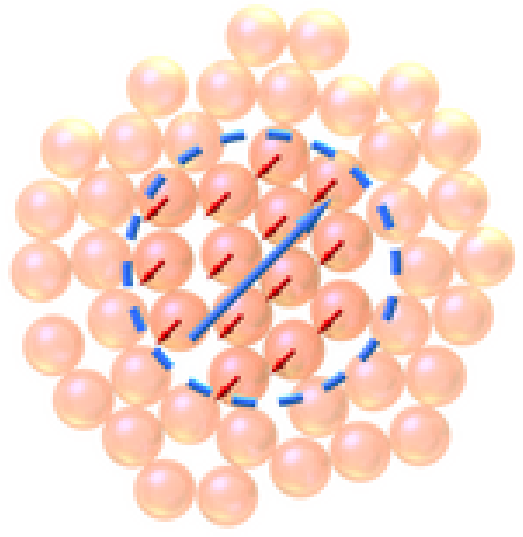}
(b)\includegraphics[width=0.25\textwidth]{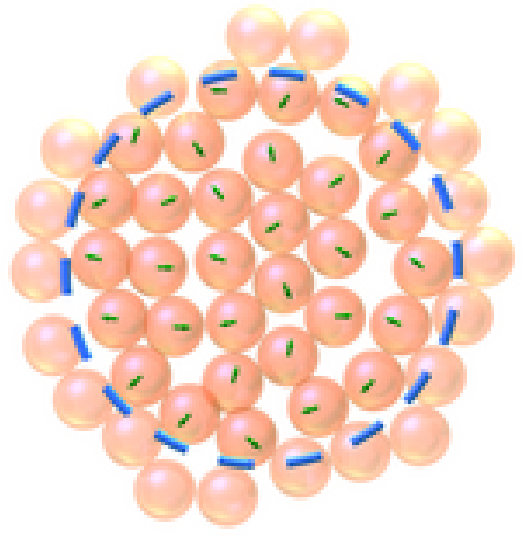}
(c)\includegraphics[width=0.25\textwidth]{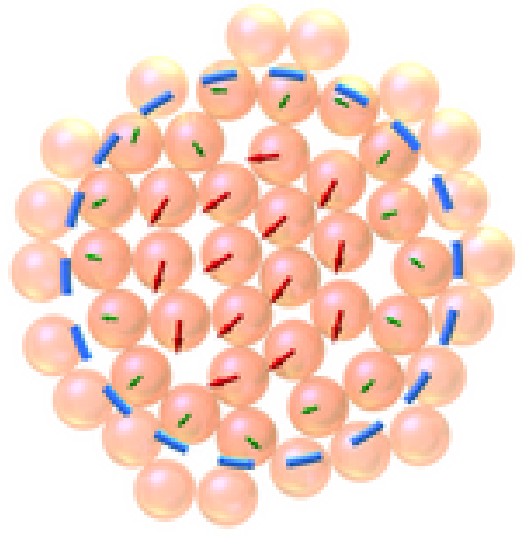}
\caption{(Color online) The mechanism for structural rearrangement in the Spot
Model. The random displacement $\vc{r}_s$ of a diffusing spot of free volume
(dashed circle) causes affected particles to move as a block by an amount
$\vc{r}_p$ (a), followed by an internal relaxation with soft-core repulsion
(b), which yields the net cooperative motion (c). (The displacements, typically
100 times smaller than the grain diameter, are exaggerated for clarity.)
\label{fig:micro}}
\end{figure*} 

A collective mechanism for random-packing dynamics has recently been proposed
to resolve this paradox and applied to granular drainage~\cite{spot-ses}. The
basic hypothesis, shown in Fig.~\ref{fig:micro}(a), is that a block of
neighboring grains makes a small, correlated downward displacement,
\begin{equation}
\Delta \vc{r}_p = - w \Delta \vc{r}_s , \label{eq:wdef}
\end{equation}
in response to the random upward displacement, $\Delta \vc{r}_s$, of a
diffusing ``spot'' of free volume. The coefficient $w$ (more generally, a
smooth function of the particle-spot separation) is set by local volume
conservation. In the simplest approximation, a spot carries a slight excess of
interstitial volume, $V_s$, spread uniformly across a sphere of radius $R_s$.
When the spot engulfs $N$ particles, each of volume $V_p$, the model predicts
$w \approx V_s/NV_p \approx \Delta\phi/\phi^2$, where $\Delta\phi$ is the local
change in volume fraction $\phi$. Allowing for some spot overlaps yields the
estimate $w \approx 10^{-2}-10^{-3}$ from the observation that $\Delta\phi/\phi
\approx 1\%$ in dense flows, which is consistent with diffusion measurements in
experiments~\cite{choi04,choiexpt} and our simulations below.  Unlike the Void
Model (which requires $w=1$), each grain's ``cage'' of nearest neighbors also
persists over long distances~\cite{choi04}; the Spot Model is able to capture
such features of drainage experiments, while remaining simple enough for
mathematical analysis, because it does not explicitly enforce packing
constraints, only the tendency of nearby particles to diffuse together.

In order to preserve valid packings, a multiscale spot algorithm has also been
suggested~\cite{spot-ses}, which we implement here for the first time. As shown
in Fig.~\ref{fig:micro}, each spot-induced block displacement (a) is followed
by a relaxation step (b), in which the affected particles and their nearest
neighbors experience a soft-core repulsion (with all other particles held
fixed). The net displacement in (c) involves a cooperative local deformation,
whose mean is roughly the block motion in (a). It is not clear {\it a priori}
that this procedure can produce realistic flowing packings, and, if so, whether
the relaxation step dominates the simple dynamics from the original model.

To answer these questions, we calibrate and test the Spot Model against
large-scale computer simulations of granular drainage, shown in
Fig.~\ref{fig:initial}. Simulations are advantageous in this case since
three-dimensional packing dynamics cannot easily be observed experimentally. We
begin by running discrete-element method (DEM) simulations, described in
section \ref{sec:demsim}. We then calibrate the free parameters in the Spot
Model by measuring various statistical quantities from the DEM simulation, as
described in \ref{sec:cali}. In section \ref{sec:spot}, we describe the
computational implementation of the Spot Model, before carrying out a detailed
comparison to DEM in section \ref{sec:res}.

\section{DEM Simulation method}\label{sec:demsim}
We employ a DEM~\cite{cundall79,landry03} to simulate $N$ frictional,
visco-elastic, spherical glass beads of diameter, $d=3\textrm{mm}$, mass $m$
under the influence of gravity $g=9.81\textrm{ms}^{-1}$. Similar to the
experiments of Refs.~\cite{choi04,choiexpt} the silo has width $50d$ and
thickness $8d$ with side walls at $x=\pm 25d$ and front and back walls at
$y=\pm 4d$, all with friction coefficient $\mu=0.5$. The initial packing is
generated by pouring $N=55,000$ particles in from a fixed height of $z=170d$
and allowing them to come to rest under gravity, filling the silo up to
$H_o\approx110d$. We also studied a taller system with $N=135,000$ generated by
pouring particles in from a height of $z=495d$, which fills the silo to $H_o
\approx 230d$. We refer to these systems by their initial height $H_o$. Drainage
is initiated by opening a circular orifice of width $8d$ centered at $x=y=0$ in
the base of the silo ($z=0$). A snapshot of all particle positions is recorded
every $2\times10^4$ time steps ($\delta t=1.75\times 10^{-6}\textrm{s}$). Once
particles drop below $z=-10d$, they are removed from the simulation.

The particles interact according to Hertzian, history dependent contact forces.
If a particle and its neighbor are separated by a distance $\vc{r}$, and they
in compression, so that $\delta=d-\abs{\vc{r}}>0$, then they experience a force
$\vc{F}=\vc{F}_n+\vc{F}_t$, where the normal and tangential components are given by
\begin{eqnarray}
\vc{F}_n&=&\sqrt{\delta/d} \left(k_n\delta\vc{n} - \frac{\gamma_n\vc{v}_n}{2}\right)\\
\vc{F}_t&=&\sqrt{\delta/d} \left(-k_t\Delta\vc{s}_t -\frac{\gamma_t\vc{v}_t}{2}\right).
\end{eqnarray}
Here, $\vc{n}=\vc{r}/\abs{\vc{r}}$. $\vc{v}_n$ and $\vc{v}_t$ are the normal
and tangential components of the relative surface velocity, and $k_{n,t}$ and
$\gamma_{n,t}$ are the elastic and viscoelastic constants, respectively.
$\Delta\vc{s}_t$ is the elastic tangential displacement between spheres,
obtained by integrating tangential relative velocities during elastic
deformation for the lifetime of the contact, and is truncated as necessary to
satisfy a local Coulomb yield criterion $\abs{\vc{F}_t}\le \mu \abs{\vc{F}_n}$.
Particle-wall interactions are treated identically, but the particle-wall
friction coefficient is set independently. For the current simulations we set
$k_t=\frac{2}{7}k_n$, and choose $k_n=2\times10^5 mg/d$. While this is
significantly less than would be realistic for glass spheres, where we expect
$k_n \sim 10^{10} mg/d$, such a spring constant would be prohibitively
computationally expensive, as the time step must have the form $\delta t
\propto k_n^{-1/2}$ for collisions to be modeled effectively. Previous
simulations have shown that increasing $k_n$ does not significantly alter
physical results \cite{landry03}. We make use of a time step of $\delta
t=1.75\times10^{-6}\textrm{s}$, and damping coefficients
$\gamma_n=\gamma_t=50\sqrt{g/d}$.

\section{Calibration of the model}\label{sec:cali}
We first look for evidence of spots in the DEM simulation and then proceed to
calibrate the model. All the calibrations are carried out for the small
$H_o=110d$ system, after which the same parameters are used for the larger
$H_o=230d$ system.

The theory predicts large numbers of spots (since many are released as each
particle exits the silo), so we seek a statistical signature of the passage of
many spots. We therefore consider the spatial correlation for velocities in
the $x$ direction, defined by
\[
C(r)=\frac{\ept{u_x(0)u_x(r)}}{\sqrt{\ept{u_x(0)^2}\ept{u_x(r)^2}}}
\]
where the expectations are taken over all pairs of velocities $(u_x(0),
u_x(r))$ of particles separated by a distance $r$ in a given test region. For a
uniform spot influence out to a cutoff radius, $R_s>d$, as shown in
Fig.~\ref{fig:micro}(a), two random particle displacements are either
identical, if they are caused by the same spot, or independent. In that case,
the spatial velocity correlation function is given by
\begin{equation}
C(r) = 
\left\{
\begin{array}{ll}
1 - \frac{3}{4}\frac{r}{R_s} + \frac{1}{16}\left(\frac{r}{R_s}\right)^3 & r<2R_s \\
0 & r\ge 2R_s \\
\end{array} \right.
\label{eq:alpha}
\end{equation}
which is the intersection volume of spheres of radius $R_s$ separated by $r$
(scaled to $1$ at $r=0$). The shape of $C(r)$ is affected by the relaxation
step in Fig.~\ref{fig:micro}(b), but the decay length is set by the spot size.

\begin{figure}
\input{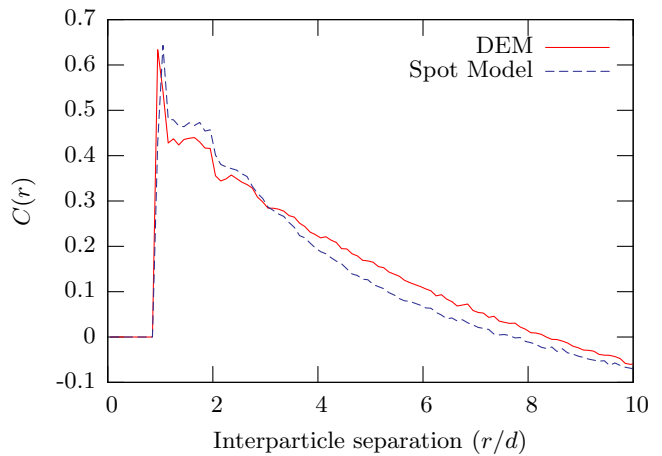}
\caption{(Color online) Comparison of velocity correlations calculated over the
time period $0.52\textrm{s}<t<1.57\textrm{s}$. Calculations are based on
particle velocity fluctuations about the mean flow in a $16d\times16d$ region
high in the center of the container. For $H_o=110d$.
\label{fig:vcorrel} }
\end{figure}

As shown in Fig.~\ref{fig:vcorrel}, we see spatial velocity correlations in the
DEM simulations at the scale of several particle diameters, consistent with the
spot hypothesis. Similar correlations have also been seen in
experiments~\cite{choi04_note} using the methods of Choi \etal~\cite{choi04},
which attests to the generality of the phenomenon, as well as the realism of
the simulations. Since the shape of $C(r)$ is not precisely that of
Eq.~(\ref{eq:alpha}), due to relaxation effects, we fit the simulation data to
a simple decay, $C(r)=\alpha e^{-r/\beta}$ with $\beta=1.87 d$. We also fit a
simple decay of the same form to Eq.~(\ref{eq:alpha}), finding $\beta=0.72
R_s$, so we infer $R_s=2.60d$ as the spot radius. Thus a grain has significant
dynamical correlations with neighbors up to three diameters away.

\begin{figure}
\input{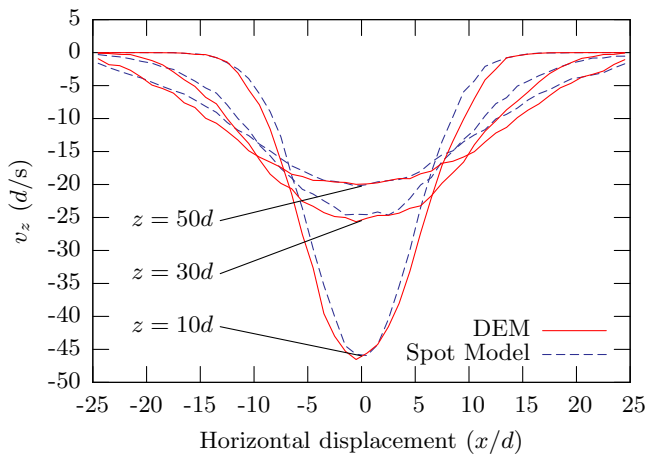}
\caption{(Color online) Comparison of the mean velocity profile, for three
different heights calculated over the time period
$4.37\textrm{s}<t<5.25\textrm{s}$ once steady flow has been established. The
Spot Model successfully predicts a Gaussian velocity profile near the orifice
and the initial spreading of the flow region with increasing height, although
the DEM flow becomes more plug-like higher in the silo.}
\label{fig:vprofile}
\end{figure}

Next, we infer the dynamics of spots, postulating independent random walks as a
first approximation. We assume that spots drift upward at a constant mean
speed, $v_s = \Delta z_s/\Delta t$, (determined below), opposite to gravity,
while undergoing random horizontal displacements of size $\Delta x_s$ in each
time step $\Delta t$. The spot diffusion length, $b_s = \Var(\Delta
x_s)/2\Delta z_s$, is obtained from the spreading of the mean flow away from
the orifice. In DEM simulations, the horizontal profile of the vertical
velocity component is well described by a Gaussian, whose variance grows
linearly with height, as shown in Fig.~\ref{fig:vprofile}. Applying linear
regression gives $\Var(u_z)=2.28zd+1.60d^2$, which implies $b_s=2.28d/2=1.14d$.
To reproduce the spot diffusion length, we chose $\Delta z_s=0.1d$ and $\Delta
x_s=0.68d$.

The typical excess volume carried by a spot can now be obtained from a single
bulk diffusion measurement. From Eq.~(\ref{eq:wdef}), the particle diffusion
length, $b_p$, is given by
\[
b_p=\frac{\Var(\Delta x_p)}{2\Delta z_p}=\frac{\Var(w \Delta x_s)}{2 w
\Delta z_s} = w b_s.
\]
We measure $b_p$ in the DEM simulation by tracking the variance of the $x$
displacements of particles that start high in the silo as a function of their
distance dropped. We find $b_p=2.86\times10^{-3}d$ and thus
$w=2.50\times10^{-3}$. During steady flow in the DEM simulation, a typical
packing fraction of particles is $57.9\%$, so a spot with radius $R_s=2.60d$
influences on average $81.7$ other particles. Thus we find that a spot carries
roughly $20\%$ of a particle volume: $V_s = 81.7 V_p/w = 0.205V_p$.

The three spot parameters so far (radius, $R_s$, diffusion length, $b_s$, and
influence factor, $w$) suffice to determine the geometrical features of a
steady flow, such as the spatial distribution of mean velocity and diffusion,
but two more are needed to introduce time dependence. The first is the mean
rate of creating spots at the orifice (for simplicity, according to a Poisson
process). In the DEM simulation, particles exit a rate of mean rate of
$4.40\times10^3\textrm{s}^{-1}$, so spots carrying a typical volume $V_s =
0.205 V_p$ should be introduced at a mean rate of $\nu_s=2.15 \times 10^4
\textrm{s}^{-1}$. The second remaining spot parameter is the vertical drift
speed, or, equivalently, the mean waiting time between spot displacements,
$\Delta t$, which can be inferred from the drop in mean packing fraction during
flow. In the DEM simulation, we find that there are initially $9,400$ particles
in the horizontal slice, $50d<z<70d$, which drops to $8,850$ during flow.
Choosing the spot waiting time to be $\Delta t = 8.68\times10^{-4}$s reproduces
this decrease in density in the spot simulation. The spot drift speed is thus
$v_s = 0.1 d /\Delta t = 115 d/\textrm{s} = 34.5\textrm{cm}/\textrm{s}$, which
is roughly ten times faster than typical particle speeds in
Fig.~\ref{fig:vprofile}.

\section{Spot model simulation}\label{sec:spot}
Having calibrated the five parameters ($R_s$, $b_s$, $w$, $\nu_s$, $v_s$), we
can test the Spot Model by carrying out drainage simulations starting from the
same static initial packing as for the DEM simulations. For efficiency, a
standard cell method (also used in the parallel DEM code) is adapted for the
spot simulations. The container is partitioned into a grid of $10 \times 3
\times N_z$ cells, each responsible for keeping track of the particles within
it, with $N_z=30$ for $H_o=110d$ and $N_z=60$ for $H_o=230d$. When a spot
moves, only the cells influenced by the spot need to be tested, and particles
are transferred between cells when necessary. Without further optimization, the
multiscale spot simulation runs over 100 times faster than the DEM simulation.

The flow is initiated as spots are introduced uniformly at random positions on
the orifice (at least $R_s$ away from the edges) at random times according to a
Poisson process of rate $\nu_s$. (The waiting time is thus an exponential
random variable of mean $\nu_s^{-1}$.) Once in the container, spots also move
at random times with a mean waiting time, $\Delta t =v_s/\Delta z_s$. Spot
displacements in the bulk are chosen randomly from four displacement vectors,
$\Delta \vc{r}_s = (\pm \Delta x_s,0,\Delta z_s), (0,\pm \Delta x_s,\Delta
z_s)$, with equal probability, so spots perform directed random walks on a
body-centered cubic lattice (with lattice parameter $2 \Delta z_s = 0.2 d$). We
make this simple choice to accelerate the simulation because more complicated,
continuously distributed and/or smaller spot displacements with the same drift
and diffusivity give very similar results. Spot centers are constrained not to
come within $d$ of a boundary, and once a spot reaches the top of the packing,
it is removed from the simulation. More realistic models for the orifice,
walls, and free surface are left for future work; here we focus on flowing
packings in the bulk.

\begin{figure*}
\begin{center}
{\large\textbf{Discrete Element Method}}\\
\vspace{0.1in}
\includegraphics[width=0.23\textwidth]{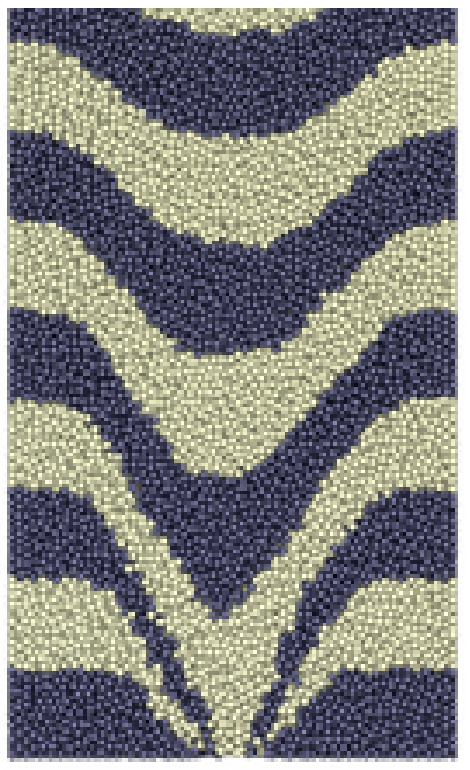}
\includegraphics[width=0.23\textwidth]{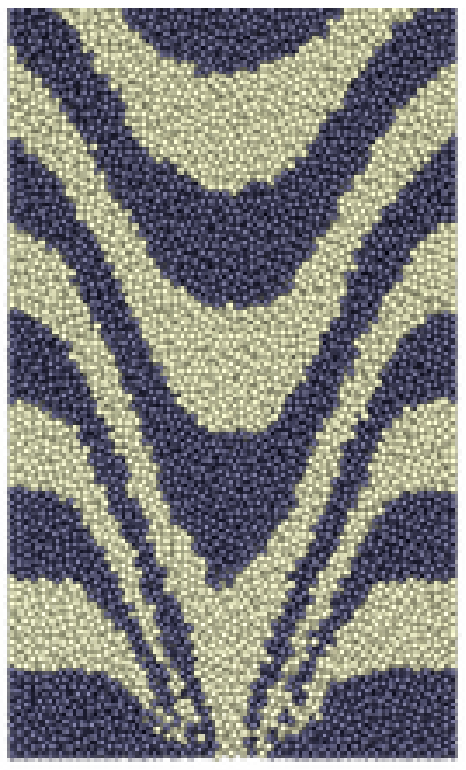}
\includegraphics[width=0.23\textwidth]{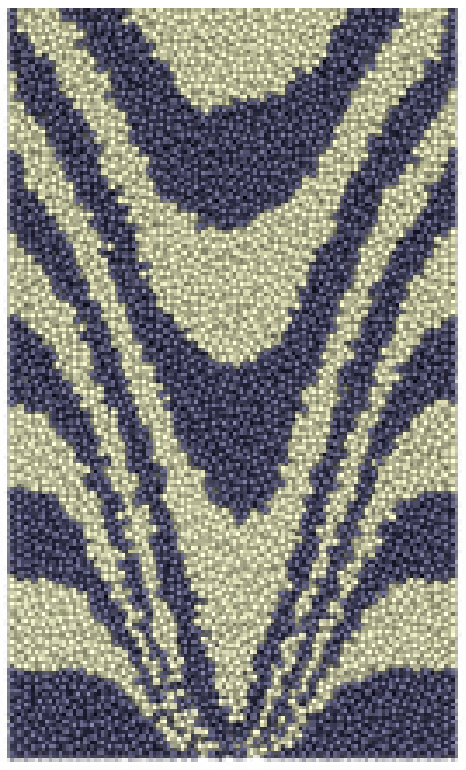}
\includegraphics[width=0.23\textwidth]{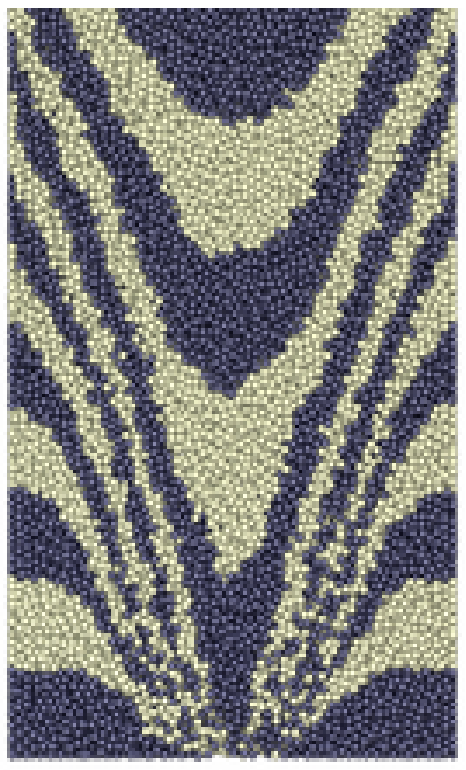}\\
\vspace{0.15in}
{\large\textbf{Spot Model}}\\
\vspace{0.1in}
\includegraphics[width=0.23\textwidth]{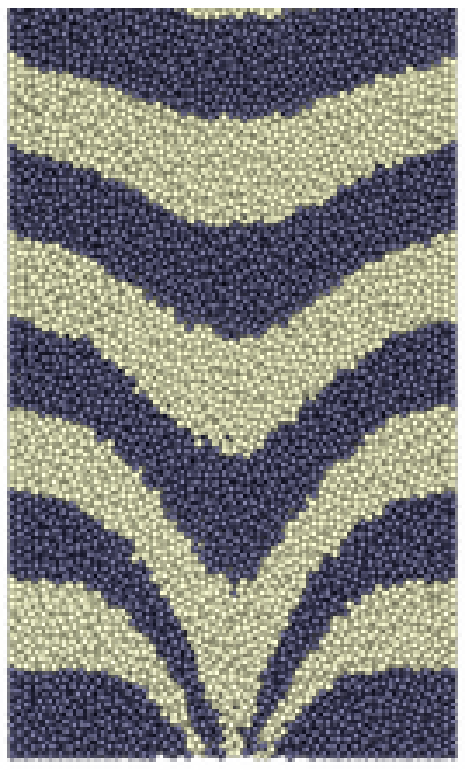}
\includegraphics[width=0.23\textwidth]{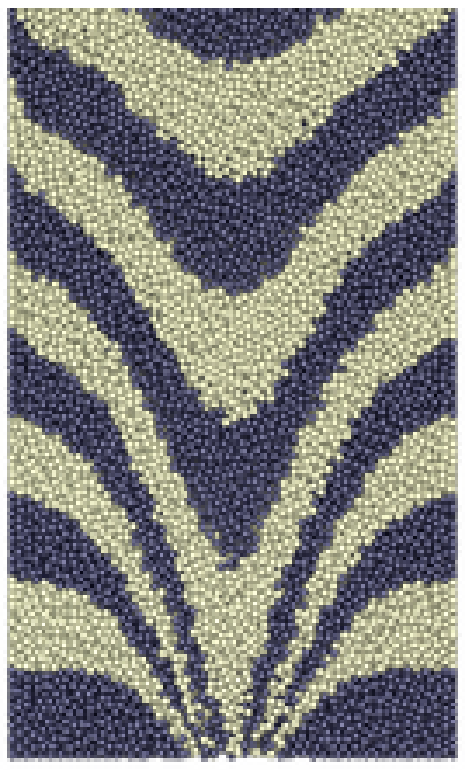}
\includegraphics[width=0.23\textwidth]{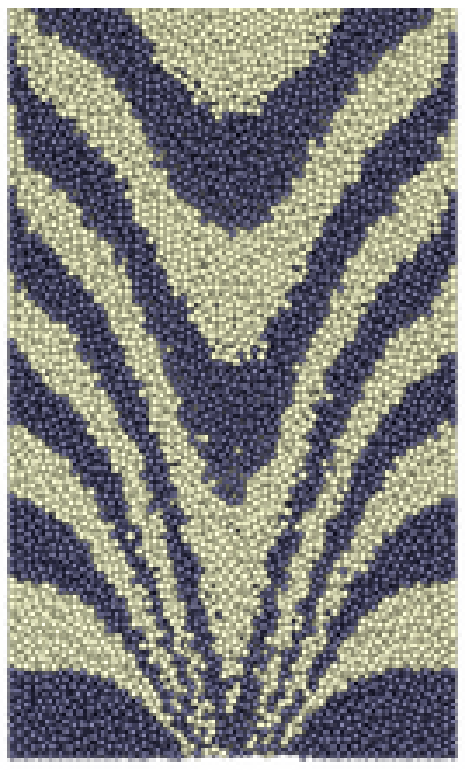}
\includegraphics[width=0.23\textwidth]{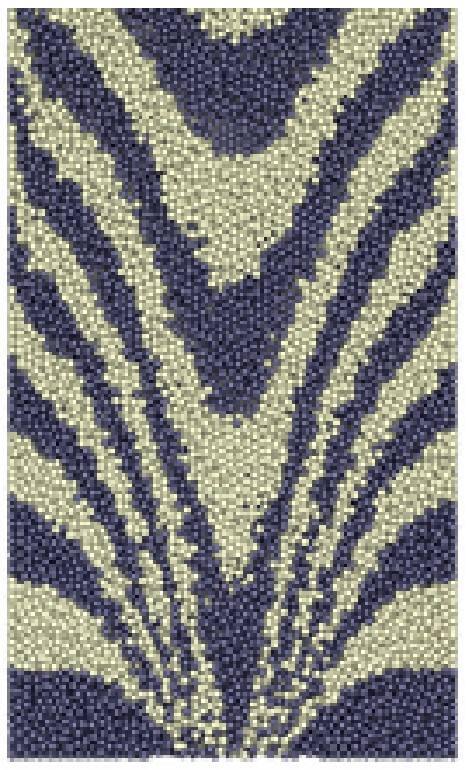}\\
\vspace{.1in}
$t=1.05$s \hspace{1.085in} 
$t=2.10$s \hspace{1.085in}
$t=3.15$s \hspace{1.085in}
$t=4.20s$ 
\caption{(Color online) Time evolution of the random packing (from left to
right) in DEM (top) and the spot simulation (bottom), for the $H_o=230d$
system, starting from the same initial state. Each image is a vertical slice
through the center of the silo near the orifice well below the free surface.}
\label{fig:snaps}
\end{center}
\end{figure*}

The particles in the simulation move passively in response to spot
displacements without any lattice constraints. Although the influence of a spot
can take a very general form~\cite{spot-ses}, the most important
aspect is its length scale, so here we choose the simplest possible model in
Eq.~(\ref{eq:wdef}), where the spot influences particles uniformly in a sphere
of radius $R_s$. As shown in Fig.~\ref{fig:micro}(a), we center the spot
influence on the midpoint of its step, which seems the most consistent with the
concept of moving interstitial volume from the initial to the final spot
position. To be precise, when a spot moves from $\vc{r}_s$ to $\vc{r}_s +
\Delta\vc{r}_s$, all particles less than $R_s$ away from $\vc{r}_s +
\Delta\vc{r}_s/2$ are displaced by $-w \Delta\vc{r}_s$.

To preserve realistic packings, we carry out a simple elastic relaxation after
each spot-induced block motion, as in Fig.~\ref{fig:micro}(b). All particles
within a radius $R_s+2d$ of the midpoint of the spot displacement exert a
soft-core repulsion on each other, if they begin to overlap. Rather than
relaxing to equilibrium or integrating Newton's laws, however, we use the
simplest possible algorithm: Each pair of particles separated by less than $d$
moves apart with identical and opposite displacements, $(d-r)\alpha$, for some
constant $\alpha>1$. Similarly, a particle within $d/2$ of a wall moves away by
a displacement, $(\frac{d}{2}-r)\alpha$. Particle positions are updated
simultaneously once all pairings are considered, but those within the shell,
$R_s+d<r<R_s+2d$, more than one diameter away from the initial block motion,
are held fixed to prevent long-range disruptions.

It turns out that, due to the cooperative nature of Spot Model, only extremely
small relaxation is required to enforce packing constraints, mainly near spot
edges where some shear occurs. Here, we choose $\alpha=0.8$ and find that the
displacements due to relaxation are typically less than 25\% of the initial
block displacement, which is at the scale of 1/10,000 of a particle diameter:
$0.25 w \Delta r_s \approx 2 \times 10^{-4} d$. Due to this tiny scale, the
details of the relaxation do not seem to be very important; we have obtained
almost indistinguishable results with $\alpha=0.6$ and $\alpha=1.0$ and also
with more complicated energy minimization schemes. As such, we do not view the
soft-core repulsion as introducing any new parameters. 

\section{Results}\label{sec:res}
The spot and DEM simulations are compared using snapshots of all particle
positions taken every $2\times10^4$ time steps. As shown in
Figure~\ref{fig:snaps}, the agreement between the two simulations is remarkably
good, considering the small number of parameters and physical assumptions in
the Spot Model. It is clear {\it a posteriori} that the relaxation step, in
spite of causing only minuscule extra displacements, manages to produce
reasonable packings during flow, while preserving the realistic description of
the mean velocity and diffusion in the basic Spot Model. Only one parameter,
$b_s$, is fitted to the mean flow, but we find that the entire velocity profile
is accurately reproduced in the lower part of the container, as shown in
Fig.~\ref{fig:vprofile}, although the flow becomes somewhat more plug-like in
DEM simulation higher in the container. Similarly, we fit $w$ to the particle
diffusion length in middle of the DEM simulation, $b_p=2.86\times10^{-3}d$,
without accounting for the elastic relaxation step, so it is reassuring that
the same measurement in the spot simulation yields a similar value,
$b_p=2.73\times10^{-3}d$.

\begin{figure}
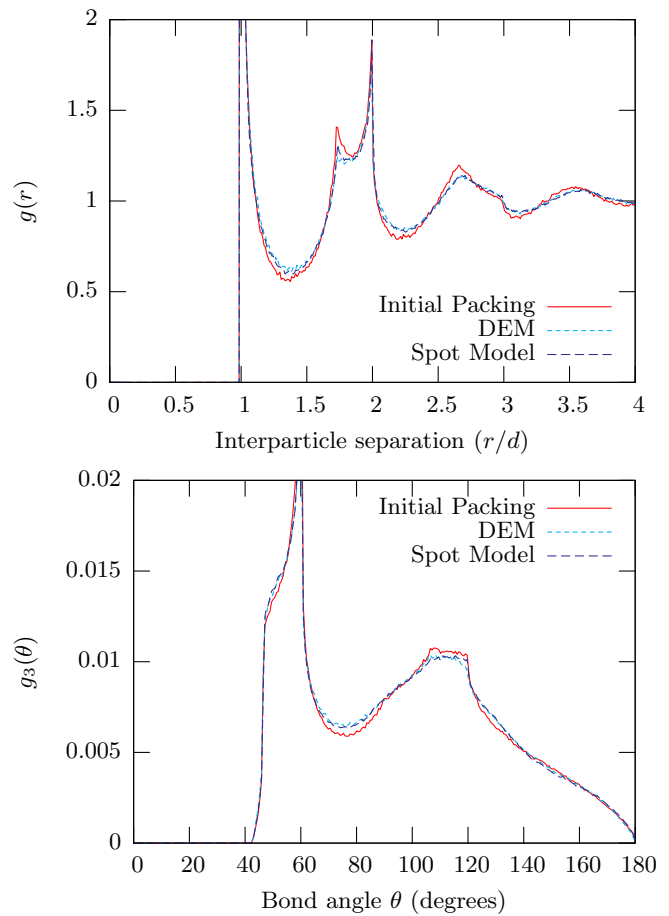

\input{finalgofr.tex}
\input{finalbond.tex}
\caption{(Color online) Comparison of radial distribution functions (top) and
bond angles (bottom) for particles in the region $-15d<x<15d$, $15d<z<45d$ for
$H_o=110d$ system. Three curves are shown on each graph, the first calculated
from the initial static packing (common between the two simulations), and the
second and third calculated for over the range
$1.04\textrm{s}<t<1.40\textrm{s}$.}
\label{fig:struct}
\end{figure}

The most surprising findings concern the agreement between the DEM and spot
simulations for various {\it microscopic} statistical quantities. First, we
consider the radial distribution function, $g(r)$, which is the distribution of
inter-particle separations, scaled to the same quantity in a ideal gas at the
same density. For dense sphere packings, the distribution begins with a large
peak near $r=d$ for particles in contact and smoothly connects smaller peaks at
typical separations of more distant neighbors, while decaying to unity.  As
shown in Fig.~\ref{fig:struct}(a), the functions $g(r)$ from the spot and DEM
simulations are nearly indistinguishable, across the entire range of neighbors
for the $H_o=110d$ system. This cannot be attributed entirely to the initial
packing because each simulation evolves independently through substantial
drainage and shearing.

Next, we consider the three-body correlation function, $g_3(\theta)$, which
gives the probability distribution for ``bond angles'' subtended by separation
vectors to first neighbors (defined by separations less than the first minimum
of $g(r)$ at $1.38d$). For sphere packings, $g_3(\theta)$ has a sharp peak at
$60^\circ$ for close-packed triangles, and another broad peak around
$110-120^\circ$ for larger crystal-like configurations. In
Fig.~\ref{fig:struct}(b), we reach the same conclusion for $g_3(\theta)$ as for
$g(r)$: The spot and DEM simulations evolve independently from the initial
packing to nearly indistinguishable steady states. 

The striking agreement between the spot and DEM simulations seems to apply not
only to structural, but also to dynamical, statistical quantities. Returning to
Fig.~\ref{fig:vcorrel}, we see that the two simulations have very similar
spatial velocity correlations. Of course, the spot size, $R_s$, in the Spot
Model (without relaxation) was fitted roughly to the scale of the correlations
in the DEM simulation, but the multiscale spot simulation also manages to
reproduce most of the fine structure of the correlation function.

\begin{figure}
\input{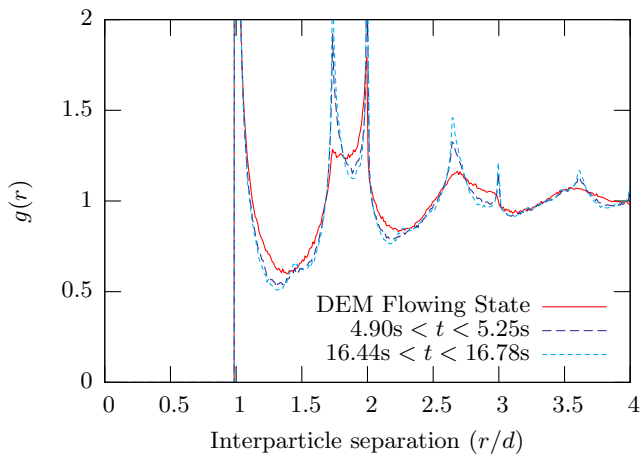}
\caption{(Color online) Evolution of the radial distribution function $g(r)$
for $H_o=230d$ in the region $-15d<x<15d$, $15d<z<45d$. The spot simulation
(dashed curves) reaches a somewhat different steady state from the DEM
simulation (solid curve), after a large amount of drainage has taken place.}
\label{fig:lterm}
\end{figure}

At much longer times, however, the random packings are no longer
indistinguishable, as a small tendency for local close-packed ordering appears
the spot simulation. As shown in Fig.~\ref{fig:lterm}, the spot simulation
develops enhanced crystal-like peaks in $g(r)$ at $r=\sqrt{3}d$, $2d$,
$\ldots$. The number of particles involved, however, is very small ($\sim 2 \%$),
and the effect seems to saturate, with no significant change between
$8\textrm{s}$ and $16\textrm{s}$.  This is consistent with even longer spot
simulations in systems with periodic boundary conditions, which reach a
similar, reproducible steady state (at the same volume fraction) from a variety
of initial conditions~\cite{jeremie05}. In all cases, the spot algorithm never
breaks down (e.g. due to jamming or instability), and unrealistic packings with
overlapping particles are never created.

The structure of the flowing steady state is fairly insensitive to various
details of the spot algorithm. For example, changing the relaxation parameter
(in the range $0.6 \leq \alpha \leq 1.0$), rescaling the spot size (by $\pm
25\%$), and using a persistent random walk (for smoother spot trajectories),
all have no appreciable effect on $g(r)$. On the other hand, decreasing the
vertical spot step size (in the range $0.025d \leq \Delta z \leq 0.1 d$) tends
to inhibit spurious local ordering and reduce the difference in $g(r)$ between
the spot and DEM simulations (e.g. measured by the $L_2$ norm). Therefore, our
spot algorithm appears to ``converge'' with decreasing time step (and
increasing computational cost), analogous to a finite-difference method,
although this merits further study.

\section{Conclusions}\label{sec:conc}
Our results suggest that {\it flowing} dense random packings have some
universal geometrical features. This would be in contrast to static dense
random packings, which suffer from ambiguities related to the degree of
randomness and definitions of jamming~\cite{torquato00,ohern03}. The similar
packing dynamics in spot and DEM simulations suggest that geometrical
constraints dominate over mechanical forces in determining structural
rearrangements, at least in granular drainage. Some form of the Spot Model may
also apply to other granular flows and perhaps even to glassy relaxation, where
localized, cooperative motion also occurs~\cite{glotzer98,weeks00}.

The Spot Model provides a simple framework for the multiscale modeling of
liquids and glasses, analogous to dislocation dynamics in crystals. Our
algorithm, which combines an efficient, ``coarse-grained'' simulation of spots
with limited, local relaxation of particles, runs over 100 times faster than
fully particle-based DEM for granular drainage. On current computers, this
means that simulating one cycle of pebble-bed reactor~\cite{talbot02} can take
hours instead of weeks~\cite{reactor}, although a general theory of spot motion
in different geometries is still lacking.  This may come from a stochastic
formulation of Mohr-Coulomb plasticity, where spots perform random walks along
slip lines of incipient failure~\cite{stoch}, which could, in principle, be
applied to different materials by changing the yield criterion.  Alternatively,
a multiscale model for supercooled molecular liquids could involve spots moving
along chains of dynamic facilitation~\cite{garrahan03,jung04}. In any case, we
have demonstrated that dense random-packing dynamics can be driven entirely by
the motion of simple, collective excitations.
\vspace{0.5cm}

\section*{Acknowledgments} 
\vspace{-0.25cm}
This work was supported by the U. S. Department of Energy (grant
DE-FG02-02ER25530) and the Norbert Weiner Research Fund and the NEC Fund at
MIT. Work at Sandia was supported by the Division of Materials Science and
Engineering, Basic Energy Sciences, Office of Science, U. S. Department of
Energy. Sandia is a multiprogram laboratory operated by Sandia Corporation, a
Lockheed Martin Company, for the U.S. Department of Energy's National Nuclear
Security Administration under contract DE-AC04-94AL85000.

\bibliography{spotmodel}

\end{document}